\documentclass{article}
\usepackage{amssymb}

\usepackage{amsmath}

\begin{document}

\title{Non-planar spin bits beyond two loops}
\author{S. Bellucci, A. Marrani \\
\textit{INFN -- Laboratori Nazionali di Frascati,}\\
\textit{Via E. Fermi 40, 00044 Frascati, Italy}}
\maketitle

\begin{abstract}
We study higher-loop orders of spin bit models underlying the non-planar
dynamics of $\mathcal{N}=4$ SYM gauge theory. In particular, we derive a
''tower'' of non-planar identities involving products of site permutation
operators. Such identities are then applied in the formulation of planarly
consistent, testable conjectures for the full non-planar, higher-loop
Hamiltonian of the $su(2)$ spin-chain.
\end{abstract}

\section{Introduction}

Large $N$ physics \cite{'tHooft:1974jz} gained considerable interest in
recent years (see \cite{Tseytlin:2004xa} for a recent review and references)
due to the AdS/CFT conjecture enlightenment \cite
{Maldacena:1998re,Gubser:1998bc} and, more recently, by considering various
limits of this correspondence (\cite{Berenstein:2002jq}-\cite
{Bellucci:2004rud}). This led to an intensive study of the anomalous
dimensions of local gauge invariant composite operators in 4-dimensional
(4-d.) $\mathcal{N}=4$ Super Yang--Mills (SYM) model \cite{Beisert:2003jj}.
A real breakthrough was the discovery of the integrability of the
Hamiltonians governing anomalous dimensions in the planar limit, $%
N\rightarrow \infty $ \cite{Minahan:2002ve,Beisert:2004ry}. These results
were extended to two and higher loops \cite{Serban:2004jf,Beisert:2003jb}.

As it is now clear, there is a one-to-one correspondence between single
trace operators in SYM theory and spin states in spin-chain models. It was
enough to consider the planar limit of SYM theory. If the non-planar
contribution is considered, the single trace sector is not conserved
anymore, and one ends up with trace splitting and joining in the operator
mixing \cite{Beisert:2002ff}. Even in this case one can still consider a
one-to-one map between local gauge invariant operators and a spin system
\cite{Bellucci:2004ru,Bellucci:2004qx}. In this case one has to introduce a
set of new degrees of freedom, beyond the spin states, which describes the
linking structure of the sites in the spin-chain. This new field takes
values in the spin bits permutation group, and introduces a new gauge degree
of freedom.

In this paper we extend the analysis of \cite{Bellucci:2005ma1} to formulate
(planarly consistent) conjectures about the explicit form of the full
non-planar $su(2)$ spin-chain Hamiltonian beyond the $2$-loop order. We will
extensively make use of some identities between (products of) permutation
operators, holding true at the non-planar level and here rigorously
obtained, clarifying their relation with the already known formulae in the
literature (see, e.g., \cite{Beisert:2003tq}).

The plan of the paper is as follows. In the next Section we introduce the
notations; then in Section 3 we obtain the above mentioned non-planar
permutational identities by simple antisymmetrization procedures. Thence, in
Section 4 we face the problem to go beyond 2-loops in non-planar spin bit
models. We formulate a set of (planarly consistent) higher-loop \emph{%
Ans\"{a}tze} in the $su(2)$ sector of \ 4-d. $\mathcal{N}=4$ SYM theory, by
using proper \emph{''deplanarizing lifts''} of site permutation operators.
Finally, in the Section 5 we draw some conclusions and perspectives for
further future developments.

Three Appendices conclude the paper: in the Appendix I we specialize the
previously obtained results on non-planar permutational identities to the $%
su(2)$ sector, particularly in relation to ultra-localization by Pauli $%
\sigma $-matrices; thence, in Appendix II the planar limit of the
permutational identities and their relation with known results are
clarified; finally, in Appendix III we consider in detail some \emph{''spin
edge-differences''} \emph{Ans\"{a}tze} for the higher-loop non-planar $su(2)$
(closed) spin-chain Hamiltonian. However, despite their intrinsic
geometrical elegance, such conjectures appear to fail, stressing the
procedure based on operator \emph{''deplanarizing lifts''} as being the only
planarly consistent approach to \emph{non-planar} $su(2)$ sector of 4-d. $%
\mathcal{N}=4$ SYM theory.

In this paper we use conventions and notations of \cite
{Bellucci:2004ru,Bellucci:2004qx,Bellucci:2005ma1}.

\section{The setup}

We consider the $su(2)$ sector of local gauge invariant SYM operators which
are generated by two holomorphic (multi)trace operators built out of two
complex SYM scalars $\phi =\phi _{5}+\mathrm{i}\phi _{6}$ and $Z=\phi _{1}+%
\mathrm{i}\phi _{2}$, of typical form
\begin{equation*}
\mathcal{O}=\mathop{\mathrm{Tr}}\nolimits(\phi Z\phi \phi Z\dots )%
\mathop{\mathrm{Tr}}\nolimits(\phi \phi \phi Z\dots )\mathop{\mathrm{Tr}}%
\nolimits(\dots )\dots
\end{equation*}

This trace can be written in the following explicit form, using a
permutation group element $\gamma \in S_{L}$:
\begin{equation}
\mathcal{O}=\phi _{i_{1}}^{a_{1}a_{\gamma _{1}}}\phi
_{i_{2}}^{a_{2}a_{\gamma _{2}}}\dots \phi _{i_{L}}^{a_{L}a_{\gamma _{L}}},
\end{equation}
where $L$ is the total number of ``letters'' $\phi _{i}=(\phi ,Z)$ in $%
\mathcal{O}$ which are numbered by a label $k=1,\dots ,L$. The permutation
group action $\gamma _{k}$ on the $k$-th label gives the next multiplier to
the $k$-th one:
\begin{equation*}
\gamma \equiv (\gamma _{1}\ \gamma _{2}\ldots \gamma _{k}\ldots \gamma
_{L})~:\quad \left(
\begin{array}{cccccc}
1 & 2 & \ldots & k & \ldots & L\cr\gamma _{1} & \gamma _{2} & \ldots &
\gamma _{k} & \ldots & \gamma _{L}
\end{array}
\ \right) \in S_{L}.
\end{equation*}

Obviously, the reshuffling of the labels $k\mapsto \sigma _{k}$ accompanied
by a conjugation of $\gamma $ with the same group element $\sigma ^{-1}\cdot
\gamma \cdot \sigma $ leaves the trace form of $\mathcal{O}$ unchanged.
Therefore, the configurations related by such a transformation should be
considered as equivalent
\begin{equation}
(\phi _{k},\gamma )\sim (\phi _{\sigma _{k}},\sigma ^{-1}\cdot \gamma \cdot
\sigma ).  \label{equiv}
\end{equation}
Now, we should map the space of such operators to the system of $L$ $su(2)$
1/2-spins (spin bits). The map is completed by associating to each bit the
spin value $\left| -1/2\right\rangle $ when we find in the respective place
the letter $\phi $, and $\left| +1/2\right\rangle $, when we find $Z$.

In perturbation theory the anomalous dimension matrix is given by
\begin{equation}
\Delta (g)=\sum_{k}H_{2k}\lambda ^{2k},  \label{dil0}
\end{equation}
with $\lambda ^{2}={\frac{g_{YM}^{2}N}{16\pi ^{2}}}$ being the 't Hooft
coupling. The coefficients in this expansion are given in terms of effective
vertices, i.e. the operators $H_{2k}$ . In principle, they can be completely
determined by an explicit evaluation of the divergencies of two-point
function $\langle \mathcal{O}(0)\mathcal{O}(x)\rangle $ Feynman amplitudes;
but such an approach is hardly feasible, specially for higher-loop orders.
The procedure based on spin bit models represents a much simpler method to
perform calculations, also at the non-planar level.

The $0$-, $1$- and $2$- loop, $su(2)$ anomalous dimension matrices are given
by the following expressions \footnote{%
Notice that, in order to obtain the correct planar limit, the expression for
$H_{4}$ (corresponding, modulo a $\frac{1}{N^{2}}$ overall factor, to
Eq.(5.5) of \cite{Beisert:2003tq}) must be properly modified, by changing $%
\frac{4}{N}$ into $\frac{2}{N}$ in front of the third term $:%
\mathop{\mathrm{Tr}}\nolimits(\left[ \phi ,Z\right] \left[ \check{\phi},%
\check{Z}\right] ):$.} \cite{Beisert:2003tq}:

\begin{eqnarray}
H_{0} &=&\mathop{\mathrm{Tr}}\nolimits(\phi \check{\phi}+Z\check{Z}), \\
H_{2} &=&-{\frac{2}{N}}:\mathop{\mathrm{Tr}}\nolimits([\phi ,Z][\check{\phi},%
\check{Z}]):, \\
H_{4} &=&\frac{1}{N^{2}}\left\{ 2:\mathop{\mathrm{Tr}}\nolimits(\left[
Z,\phi \right] \left[ \check{Z},\left[ Z,\left[ \check{Z},\check{\phi}\right]
\right] \right] ):\right. +2:\mathop{\mathrm{Tr}}\nolimits(\left[ Z,\phi %
\right] \left[ \check{\phi},\left[ \phi ,\left[ \check{Z},\check{\phi}\right]
\right] \right] ):+  \notag \\
&&\qquad \left. +2N:\mathop{\mathrm{Tr}}\nolimits(\left[ \phi ,Z\right] %
\left[ \check{\phi},\check{Z}\right] ):\right\} ,
\end{eqnarray}
where the checked letters $\check{\phi}$ and $\check{Z}$ correspond to
derivatives with respect to the matrix elements
\begin{equation}
\check{Z}_{ij}=\frac{\partial }{\partial Z^{ji}},\qquad \check{\phi}_{ij}=%
\frac{\partial }{\partial \phi ^{ji}}
\end{equation}
and colons denote the ordering, in which all checked letters in the group
are assumed to stay on the right of the unchecked ones.

In order to find the ``pull back'' of the Hamiltonian \eqref{dil0} to the
spin description, one has to apply it on a (multi)trace operator
corresponding to the spin bit state $\left|s,\gamma\right\rangle$ and map
the result back to the corresponding spin bit state. This can be done
term-by-term in the perturbation theory expansion series.

The $1$-loop Hamiltonian was found earlier \cite
{Bellucci:2004ru,Bellucci:2004qx} and reads\footnote{%
Because of the periodic boundary conditions (p.b.c.) assumed for the closed
spin-chain, the subscript site indices may equivalently range in $\left\{
1,...,L\right\} $ or in $Z_{L}$, i.e. in the integer numeric field with
period $L$.
\par
$H_{k_{1}k_{2}}\equiv 1-P_{k_{1}k_{2}}$ is the two-site, planar $su(2)$ $1$%
-loop spin bit Hamiltonian; it is nothing but twice the site index
antisymmetrizer $\frac{1}{2}(1-P_{k_{1}k_{2}})$, and therefore it makes the
constraint $k_{1}\neq k_{2}$ redundant.}

\begin{eqnarray}
\ H_{2} &=&{\textstyle\frac{1}{N}}\sum_{k_{1},k_{2}=1,\left( k_{1}\neq
k_{2}\right) }^{L}H_{k_{1}k_{2}}\left( \Sigma _{\gamma _{k_{1}}k_{2}}+\Sigma
_{k_{1}\gamma _{k_{2}}}-\Sigma _{k_{1}k_{2}}-\Sigma _{\gamma _{k_{1}}\gamma
_{k_{2}}}\right) =  \notag \\
&=&{\textstyle\frac{1}{N}}\sum_{k_{1},k_{2}=1}^{L}(P_{k_{1}k_{2}}-1)\left(
\Sigma _{k_{1}k_{2}}+\Sigma _{\gamma _{k_{1}}\gamma _{k_{2}}}-\Sigma
_{\gamma _{k_{1}}k_{2}}-\Sigma _{k_{1}\gamma _{k_{2}}}\right) ,  \label{d2}
\end{eqnarray}
where the site index permutation and chain ``twist'' operators are
respectively defined in the following way $(k_{1},k_{2}=1,...,L)$:

\begin{gather}
P_{k_{1}k_{2}}\,\left| \{\dots A_{k_{1}}\dots A_{k_{2}}\dots \}\right\rangle
=\left| \{\dots A_{k_{2}}\dots A_{k_{1}}\dots \}\right\rangle
,A_{k_{2}},A_{k_{1}}=\phi ,Z  \label{sigma} \\
\notag \\
\Sigma _{k_{1}k_{2}}\left| \gamma \right\rangle =
\begin{cases}
\left| \gamma \,\sigma _{k_{1}k_{2}}\right\rangle & \text{if }k_{1}\neq k_{2}
\\
N\left| \gamma \right\rangle , & k_{1}=k_{2}.
\end{cases}
\label{sigma2}
\end{gather}

$\Sigma _{k_{1}k_{2}}$ acts as a chain splitting and joining operator. The
factor $N$ in the case $k_{1}=k_{2}$ in Eq. \eqref{sigma} appears because
the splitting of a trace at the same place leads to a chain of length zero,
whose corresponding trace is ${\mathop{\mathrm{Tr}}\nolimits}\,1=N$. It is
important to note that the operator $\Sigma _{k_{1}k_{2}}$ acts only on the
linking variable, while the two-site $su(2)$ $1$-loop spin bit Hamiltonian $%
H_{k_{1}k_{2}}=(1-P_{k_{1}k_{2}})$ acts on the spin space. Therefore, the
two operators commute.

\section{Identities of permutation operators at the non-planar level}

Given three site indices $k_{1},k_{2},k_{3}$ all different from each other
(this assumption will be denoted, here and further below, by the notation $%
k_{1}\neq k_{2}\neq k_{3}$), the following identity holds ($\left(
k_{1},k_{2},k_{3}\right) \in \left\{ 1,...,L\right\} ^{3},$ $k_{1}\neq
k_{2}\neq k_{3}$):
\begin{equation}
P_{k_{1}k_{2}}P_{k_{1}k_{3}}=P_{k_{2}k_{3}}P_{k_{1}k_{2}}=P_{k_{1}k_{3}}P_{k_{2}k_{3}},
\label{1st-id}
\end{equation}
expressing the redundancy in the product of two permutation operators with
one repeated index ($L$ may be here considered as the total number of
spin-chain sites, and also the spin-chain length, if a unit distance between
adiacent sites is assumed).

Notice that, differently from the identities reported below, (\ref{1st-id})
is always true, even out of the $su(2)$ symmetry operatorial sector of the
considered 4-d. $\mathcal{N}=4$ SYM theory.

Many other identities, holding at the non-planar level, may be obtained by
simply antisymmetrizing tensors of rank $\geqslant M$ in operatorial sectors
with spin(-site indice)s\footnote{%
Round brackets may equivalently be put or not, because of
\begin{equation*}
P_{k_{1}k_{2}}=P_{\widehat{S}_{k_{1}}\widehat{S}_{k_{2}}},
\end{equation*}
$\widehat{S}_{k_{1}}$ and $\widehat{S}_{k_{2}}$ denoting the set of spin
operators on the $k_{1}$-th and $k_{2}$-th spin-chain sites, respectively.}
having range(s) with cardinality $\leqslant M-1$ on each site. In the
following treatment, we will mainly focus on the $su(2)$ symmetry sector,
with a spin $s=\frac{1}{2}$ irreducible representation on each site, whence $%
M-1=2s+1\Leftrightarrow M=3$.

Thus, from the antisymmetrization of a 3-index tensor we get ($1$ denotes,
here and further below, the permutational identity operator and, as before, $%
\left( k_{1},k_{2},k_{3}\right) \in \left\{ 1,...,L\right\} ^{3},$ $%
k_{1}\neq k_{2}\neq k_{3}$):
\begin{equation}
1-P_{k_{1}k_{2}}-P_{k_{1}k_{3}}-P_{k_{2}k_{3}}+P_{k_{1}k_{2}}P_{k_{2}k_{3}}+P_{k_{2}k_{3}}P_{k_{1}k_{2}}=0,
\label{2nd-id}
\end{equation}
holding true in general in each symmetry sector with spin(-site indice)s
ranging in sets with cardinality $\leqslant 2$.

Analogously, antisymmetrizing a 4-index tensor, we get ($\left(
k_{1},k_{2},k_{3},k_{4}\right) \in \left\{ 1,...,L\right\} ^{4},$ $k_{1}\neq
k_{2}\neq k_{3}\neq k_{4},$):
\begin{equation}
\begin{array}{l}
1-P_{k_{1}k_{2}}-P_{k_{1}k_{3}}-P_{k_{1}k_{4}}-P_{k_{2}k_{3}}-P_{k_{2}k_{4}}-P_{k_{3}k_{4}}+
\\
\text{ \ }%
+P_{k_{3}k_{4}}P_{k_{2}k_{3}}+P_{k_{3}k_{4}}P_{k_{2}k_{4}}+P_{k_{3}k_{4}}P_{k_{1}k_{2}}+P_{k_{3}k_{4}}P_{k_{1}k_{3}}+P_{k_{3}k_{4}}P_{k_{1}k_{4}}+
\\
\text{ \ }%
+P_{k_{2}k_{3}}P_{k_{1}k_{2}}+P_{k_{2}k_{3}}P_{k_{1}k_{3}}+P_{k_{2}k_{3}}P_{k_{1}k_{4}}+
\\
\text{ \ }%
+P_{k_{2}k_{4}}P_{k_{1}k_{2}}+P_{k_{2}k_{4}}P_{k_{1}k_{3}}+P_{k_{2}k_{4}}P_{k_{1}k_{4}}+
\\
\text{ \ }%
-P_{k_{3}k_{4}}P_{k_{2}k_{3}}P_{k_{1}k_{2}}+P_{k_{3}k_{4}}P_{k_{2}k_{3}}P_{k_{1}k_{3}}+P_{k_{3}k_{4}}P_{k_{2}k_{3}}P_{k_{1}k_{4}}+
\\
\text{ \ }%
-P_{k_{3}k_{4}}P_{k_{2}k_{4}}P_{k_{1}k_{2}}+P_{k_{3}k_{4}}P_{k_{2}k_{4}}P_{k_{1}k_{3}}+P_{k_{3}k_{4}}P_{k_{2}k_{4}}P_{k_{1}k_{4}}=0,
\end{array}
\label{3rd-id}
\end{equation}
which is true in general, in each symmetry sector with spin(-site indice)s
ranging in sets with cardinality $\leqslant 3$.

Looking at (\ref{2nd-id}) and (\ref{3rd-id}) and using Eq. (\ref{1st-id}),
it can be shown by explicit calculations that such identities have,
respectively, the following structure:
\begin{equation}
\sum_{\pi \in S_{3}}\left( -\right) ^{\sigma _{\pi }}P_{\pi }=0,
\end{equation}
\begin{equation}
\sum_{\pi \in S_{4}}\left( -\right) ^{\sigma _{\pi }}P_{\pi }=0;
\end{equation}
Here, as previously defined, $S_{3}$ and $S_{4}$ are the permutation groups
of 3 and 4 different spin-chain sites, and $\pi $ stands for one element of
such groups; its realization in terms of (a product of) spin-chain site pair
permutation operators $P$'s is denoted with $P_{\pi }$. $\sigma _{\pi }$ is
defined \emph{''length of permutation }$\pi $\emph{''}, and depends on the
realization through $P$'s: $\pi $ will have $\sigma _{\pi }=0,1,2,3,...$ if
it is proportional to identity operator, linear, quadratic, cubic and so on
in $P$'s, respectively. Finally, $\left( -\right) ^{\sigma _{\pi }}$ is
called \emph{''parity''} of the $\sigma _{\pi }$-lengthed permutation $\pi $%
. Notice that, because of the cardinality of discrete groups $S_{3}$ and $%
S_{4}$ is respectively $6$ and $24$, Eqs. (\ref{2nd-id}) and (\ref{3rd-id})
do contain the right number of terms in their l.h.s.s.

Therefore, it is possible to argue the general structure of the identities
on permutation operators $P$'s, arising from the antisymmetrization
procedure of $M(\geqslant 3)$-index tensors in symmetry sectors of 4-d. $%
\mathcal{N}=4$ SYM theory having spin(-site indice)s ranging in sets with
cardinality $\leqslant M-1$:
\begin{equation}
\sum_{\pi \in S_{M}}\left( -\right) ^{\sigma _{\pi }}P_{\pi }=0,
\label{gen-struct}
\end{equation}
with the l.h.s. containing $M!$ independent permutations (corresponding to
all the elements of the group $S_{M}$), realized through (product of) site
pair permutation operators $P$'s, acting on a set $\left\{
k_{1},k_{2},...,k_{M}\right\} $ of $M$ different spin-chain site indices.

Notice that all identities obtainable by varying $M\geqslant 3$ in (\ref
{gen-struct}) are \emph{independent}, i.e. they cannot be obtained one from
the other by applying one or more permutation operators. this is trivially
evident by comparing (\ref{2nd-id}) with (\ref{3rd-id}), because the
application of $P$'s carrying new spin(-site indice)s does not allow to
obtain the correct number of independent higher-order products of $P$'s.

At this point, it would be very interesting, and we leave it for further
future investigations, to see if general recursive (possibly algorithmical)
formulae for (\ref{gen-struct}) as function of $M\geqslant 3$ may exist and
be given; if so, probably they quite intriguingly link the identities among
(products of) permutation operators $P$'s with the \emph{random graph theory}
in a discrete set of sites. An interesting task would be to study such a
connection, depending on the imposed \emph{boundary conditions} (periodic in
our case), intimely related to the kind of string spin bits yield a \emph{%
dynamical discretization} of.

\section{\emph{''Deplanarizing operator lifts''} and planarly consistent
higher-loop \emph{Ans\"{a}tze} in the $su(2)$ sector of 4-d. $\mathcal{N}=4$
SYM theory}

Let us now consider the \emph{non-planar} spin-chain Hamiltonian in the $%
su(2)$ sector of the 4-d. $\mathcal{N}=4$ SYM theory.

As already reported in Eq. (\ref{d2}), at the $1$-loop level in perturbation
theory it reads \cite{Bellucci:2004ru,Bellucci:2004qx} (for simplicity's
sake, we omit to ''check'' the symbols denoting spin operators, as instead
rigorously done in Appendix I)
\begin{eqnarray}
H_{2} &=&\frac{2}{N}\sum_{\left( k_{1},k_{2}\right) \in Z_{L}^{2}}\left(
1-P_{k_{1}k_{2}}\right) \Sigma _{k_{1}\gamma _{k_{2}}}=\frac{2}{N}%
\sum_{\left( k_{1},k_{2}\right) \in Z_{L}^{2}}H_{k_{1}k_{2}}\Sigma
_{k_{1}\gamma _{k_{2}}}=  \notag \\
&=&\frac{2}{N}\sum_{\left( k_{1},k_{2}\right) \in Z_{L}^{2},\text{ \ }%
k_{1}\neq k_{2}}H_{k_{1}k_{2}}\Sigma _{k_{1}\gamma _{k_{2}}}=  \notag \\
&=&\frac{2}{N}\sum_{\left( k_{1},k_{2}\right) \in Z_{L}^{2},\text{ \ }%
k_{1}\neq k_{2}}\left[ \left( \overrightarrow{S_{k_{1}}}-\overrightarrow{%
S_{k_{2}}}\right) ^{2}-1\right] \Sigma _{k_{1}\gamma _{k_{2}}},  \label{34}
\end{eqnarray}
where $H_{k_{1}k_{2}}$ is the previously defined two-site, planar, 1-loop $%
su(2)$ spin chain Hamiltonian (coinciding with twice the site
antisymmetrizer), and in the last passage we used Eqs. (\ref{16}) and (\ref
{18}), implying that
\begin{equation}
P_{k_{1}k_{2}}-1=1-\left( \overrightarrow{S_{k_{1}}}-\overrightarrow{%
S_{k_{2}}}\right) ^{2}-\delta _{k_{1}k_{2}},\text{ \ \ }\forall \left(
k_{1},k_{2}\right) \in Z_{L}^{2}.
\end{equation}
Notice that $H_{2}$ is completely symmetric under the exchange $\left(
k_{1},k_{2}\right) \leftrightarrow \left( k_{2},k_{1}\right) $, as it has to
be from the site-index structure obtained from the linking part $\Sigma
_{k_{1}\gamma _{k_{2}}}$ of the Hamiltonian.

Analogously, at the $2$-loop level, in \cite{Bellucci:2005ma1} the following
form for the non-planar $su(2)$ spin-chain Hamiltonian was obtained:
\begin{eqnarray}
H_{4} &=&\frac{2}{N^{2}}\sum_{\left( k_{1},k_{2},k_{3}\right) \in
Z_{L}^{3}}\left( 2P_{k_{1}k_{2}}-P_{k_{1}k_{3}}+2P_{k_{2}k_{3}}-3\right)
\Sigma _{k_{1}\gamma _{k_{2}}}\Sigma _{k_{2}\gamma _{k_{3}}}=  \notag \\
&=&-\frac{2}{N^{2}}\sum_{\left( k_{1},k_{2},k_{3}\right) \in Z_{L}^{3},\text{
\ }k_{1}\neq k_{2}\neq k_{3}}\left(
2H_{k_{1}k_{2}}-H_{k_{1}k_{3}}+2H_{k_{2}k_{3}}\right) \Sigma _{k_{1}\gamma
_{k_{2}}}\Sigma _{k_{2}\gamma _{k_{3}}}=  \notag \\
&=&\frac{2}{N^{2}}\sum_{\left( k_{1},k_{2},k_{3}\right) \in Z_{L}^{3},\text{
\ }k_{1}\neq k_{2}\neq k_{3}}\left( \overrightarrow{S_{k_{1}}}-2%
\overrightarrow{S_{k_{2}}}+\overrightarrow{S_{k_{3}}}\right) ^{2}\Sigma
_{k_{1}\gamma _{k_{2}}}\Sigma _{k_{2}\gamma _{k_{3}}}.  \label{36}
\end{eqnarray}
Notice that $H_{4}$ is completely symmetric under the exchange $\left(
k_{1},k_{2},k_{3}\right) \leftrightarrow \left( k_{3},k_{2},k_{1}\right) $,
as it has to be from the site-index structure obtained from the linking part
$\Sigma _{k_{1}\gamma _{k_{2}}}\Sigma _{k_{2}\gamma _{k_{3}}}$ of the
Hamiltonian.

By comparing Eqs. (\ref{34}) and (\ref{36}), it appears reasonable to
formulate an elegant and simply-meaning set of conjectures, named \emph{%
''spin edge-differences''} \emph{Ans\"{a}tze}, for the explicit form of the $%
n$-loop, non-planar $su(2)$ spin-chain Hamiltonian $H_{2n}$; they are
treated in some detail in Appendix III. Such \emph{Ans\"{a}tze }%
unfortunately appear to fail; therefore, if we want to go beyond the $2$%
-loop level in (non-planar) spin bit models, we have to find other ways to
formulate consistent \emph{Ans\"{a}tze} for the explicit form of $H_{2n}$,
for $n\geqslant 3$.

To achieve this, let us start introducing the compact notation \cite
{Beisert:2003tq}
\begin{equation}
\left\{ n_{1},n_{2},...\right\} \equiv
\sum_{k=1}^{L}P_{k+n_{1},k+n_{1}+1}P_{k+n_{2},k+n_{2}+1}...,\text{ \ \ \ \ }%
n_{1},n_{2},...\in Z,  \label{notation-planar}
\end{equation}
and let us report the currently known planar $su(2)$ spin-chain Hamiltonians
\cite{Beisert:2003tq}, \cite{Beisert:2003hl}:
\begin{eqnarray}
\text{tree level} &:&\text{ \ }H_{0,planar}=\{\},  \label{D0} \\
1\text{-loop level} &:&\text{ }H_{2,planar}=2\{\}-2\{0\},  \label{D2} \\
2\text{-loop level} &:&\text{ }H_{4,planar}=-8\{\}+12\{0\}-2\left( \left\{
1,0\right\} +\left\{ 0,1\right\} \right) ,  \label{D4}
\end{eqnarray}
\begin{eqnarray}
3\text{-loop level} &:&H_{6,planar}=  \notag \\
&=&60\{\}-104\{0\}+24\left( \left\{ 1,0\right\} +\left\{ 0,1\right\} \right)
+  \notag \\
&&+4\{0,2\}-4\left( \left\{ 0,1,2\right\} +\left\{ 2,1,0\right\} \right) ,
\label{D6}
\end{eqnarray}
\begin{eqnarray}
4\text{-loop level} &:&H_{8,planar}\left( \beta \right) =  \notag \\
&=&-560\{\}+(1036+4\beta )\{0\}+(-266-4\beta )(\{0,1\}+\{1,0\})+  \notag \\
&&+(-66-2\beta )\{0,2\}-4\{0,3\}+  \notag \\
&&+4(\{0,1,3\}+\{0,2,3\}+\{0,3,2\}+\{1,0,3\})+  \notag \\
&&+(78+2\beta )(\{0,1,2\}+\{2,1,0\})+(-18+2\beta )(\{0,2,1\}+\{1,0,2\})+
\notag \\
&&+(1-\beta )(\{0,1,3,2\}+\{0,3,2,1\}+\{1,0,2,3\}+\{2,1,0,3\})+  \notag \\
&&+(6-2\beta )\{1,0,2,1\}+2\beta (\{0,2,1,3\}+\{1,0,3,2\})+  \notag \\
&&-10(\{0,1,2,3\}+\{3,2,1,0\}),  \label{D8}
\end{eqnarray}
where in the last $4$-loop expression $\beta $ is an undetermined real
parameter; it is \emph{unphysical}, because it corresponds to rotations of
the space of the states of the $su(2)$ spin-chain system, and therefore it
changes the (unphysical) eigenstates, but not the (physical) eingeinvalues
\cite{Beisert:2004di}.

As already mentioned in \cite{Bellucci:2005ma1}, from the definition (\ref
{sigma2}) of the \emph{''twist''} operator $\Sigma _{kl}$ the following
decomposition holds:
\begin{equation}
\Sigma _{kl}=N\delta _{kl}+\left( 1-\delta _{kl}\right) \widetilde{\Sigma }%
_{kl},
\end{equation}
where $\widetilde{\Sigma }_{kl}$ is the ''real'' chain splitting and joining
operator, spoiled of its degeneracy in the case of coinciding sites. Whence
the planar limit $N\rightarrow \infty $ affects just the ''twist'' operator,
and it does in the following way:
\begin{equation}
\lim_{N\rightarrow \infty }\frac{1}{N}\Sigma _{kl}=\delta _{kl}.
\end{equation}
Consequently, using the fact that, from Eq. (\ref{equiv}) and from the
periodic boundary conditions for the closed spin-chain, the following
identity trivially holds ($\forall \gamma \in S_{L}$):
\begin{equation}
\sum_{k\in Z_{L}}P_{k\gamma _{k}}=\sum_{k\in Z_{L}}P_{\gamma _{k}\gamma
_{k}^{2}},  \label{51}
\end{equation}
and passing to a \emph{canonical} form (in which $\gamma _{k}=k+1$, $\gamma
_{k}^{2}=k+2$, and so on), it is easy to check that the planar limits of
Eqs. (\ref{34}) and (\ref{36}) perfectly match (\ref{D2}) and (\ref{D4}),
respectively.

Whence, comparing Eqs. (\ref{D2}) and (\ref{D4}) with their full non-planar
versions, respectively given by Eqs. (\ref{34}) and (\ref{36}), it is easy
to see that they are related in the following simple way: each term of the
full non-planar spin part, once one identifies (without loss of generality) $%
k_{2}=\gamma _{k}$ and $k_{3}=\gamma _{k}^{2}$, gives rise, by the use of (%
\ref{51}), to the corresponding planar term of the spin part, with the
correct coefficients.

The generalization of such a procedure to the $3$- and $4$-loop levels gives
a (planarly consistent) way to formulate higher-loop \emph{Ans\"{a}tze} for
the spin part of the non-planar, $su(2)$ spin-chain Hamiltonian. For what
concerns the linking variable part, i.e. the $\Sigma $'s, the same
conjectures used in the \emph{''spin edge-differences''} approach explained
in Appendix III will be assumed; this amounts to identifying the linking
variable part of the $n$-loop non-planar Hamiltonian with the ($\gamma $%
-dependent) \emph{''splitting and joining chain operator of order }$n$\emph{%
''}, defined as
\begin{equation}
\Sigma _{k_{0}k_{1}...k_{n}}\left( \gamma \right) \equiv \Sigma
_{k_{0}\gamma _{k_{1}}}\Sigma _{k_{1}\gamma _{k_{2}}}...\Sigma
_{k_{n-2}\gamma _{k_{n-1}}}\Sigma _{k_{n-1}\gamma _{k_{n}}},\text{ \ \ \ }%
\gamma \in S_{L}.  \label{38bis}
\end{equation}
Such a definition necessarily implies the invariance of the spin part under
the \emph{''site-index inversion''}
\begin{equation}
\left( k_{0},k_{1},k_{2},...,k_{n-1},k_{n}\right) \leftrightarrow \left(
k_{n},k_{n-1},k_{n-2},...,k_{1},k_{0}\right) .  \label{exchange}
\end{equation}
It is worth remarking that such an assumption, beside being well-motivated
from the knowledge of the first loop orders, does not take into account
possible contributions from higher-loop Feynman's diagrammatics. Indeed,
\emph{a priori} one should consider that the linking variable part of the $m$%
-loop non-planar Hamiltonian should include all $\Sigma
_{k_{0}k_{1}...k_{n}}\left( \gamma \right) $, with $1\leqslant n\leqslant m$%
; this is due to the fact that usually all structures arising at a given
loop order reappear at higher orders, because an insertion of an internal
loop into a lower-order diagram produces structurally equivalent
contribution to the Hamiltonian. Nevertheless, as it is evident by comparing
Eqs. (\ref{34}) and (\ref{36}), this does \emph{not} happen at $1$- and $2$%
-loops, where the above-formulated conjecture based on the ($\gamma $%
-dependent) \emph{''splitting and joining chain operator of order }$n$\emph{%
''} defined in Eq. (\ref{38bis}) perfectly matches the already known (and
independently obtained) non-planar results; therefore we assume it to hold
true also at higher-loop orders.

Thus, we have to start from the planar, $3$- and $4$-loop expressions of the
$su(2)$ spin-chain Hamiltonian, respectively given by Eqs. (\ref{D6}) and (%
\ref{D8}), and consider \emph{all} possible products of $P$'s that, in the
planar limit, would give the considered planar permutational term. All such
non-planar permutational terms will come with free (real) coefficients,
constrained by two request:

$i)$ their sum must give the right numerical known coefficient of the
considered planar permutational term;

$ii)$ they must make the spin part of the non-planar Hamiltonian completely
symmetric under the proper exchange of site indices (\ref{exchange}), as
requested by the site index structure determined by the linking part.

Then, making the (non-reductive) conventional site identifications $%
k_{2}=\gamma _{k}$, $k_{3}=\gamma _{k}^{2}$ and so on, and adding the
linking variable using the definition (\ref{38bis}), a complete \emph{Ansatz}
for the full non-planar, $su(2)$ spin-chain Hamiltonian at the considered
loop order is obtained\footnote{%
It should be noticed that we assume that (eventually rather structurally
complicated) non-planar permutational terms, such that their planar limit is
zero, do \emph{not} exist; indeed, for the time being, their existence may
not be guessed by an inferring approach starting from the planar level, such
as the one adopted in this paper.}.

Let us consider an explicit example of the described method, in order to
build a consistent \emph{Ansatz} for the expression of the full non-planar, $%
3$-loop $su(2)$ spin-chain Hamiltonian.

We start from the expression of $H_{6,planar}$ given by (\ref{D6}). By
conventionally identifying (without loss of generality) the spin chain-chain
site indices in the following way:
\begin{equation}
k_{1}\equiv k,\text{ \ }k_{2}\equiv \gamma _{k},\text{ \ }k_{3}\equiv \gamma
_{k}^{2},\text{ \ }k_{4}:=\gamma _{k}^{3},
\end{equation}
we therefore have to find \emph{all} possible non-planar permutational terms
giving rise, in the planar limit $N\rightarrow \infty $, to each of the
permutational terms of $H_{6,planar}$ given by Eq. (\ref{D6}). We have that:

$i)$ the planar term $P_{k\gamma _{k}}$ receives three kind of contributions
from the non-planar level, respectively from $P_{k\gamma
_{k}}=P_{k_{1}k_{2}} $, $P_{\gamma _{k}\gamma _{k}^{2}}=P_{k_{2}k_{3}}$ and $%
P_{\gamma _{k}^{2}\gamma _{k}^{3}}=P_{k_{3}k_{4}}$, whence the proper \emph{%
''deplanarizing operator lift''} of $P_{k\gamma _{k}}$ reads ($\zeta
_{1},\zeta _{2}\in R$)
\begin{equation}
-26P_{k\gamma _{k}}\rightarrow \zeta _{1}P_{k_{1}k_{2}}+\zeta
_{2}P_{k_{2}k_{3}}-(26+\zeta _{1}+\zeta _{2})P_{k_{3}k_{4}};
\end{equation}

$ii)$ the planar-level product $P_{k\gamma _{k}}P_{\gamma _{k}\gamma
_{k}^{2}}$ instead receives contribution just from two non-planar terms,
i.e. $P_{k\gamma _{k}}P_{\gamma _{k}\gamma
_{k}^{2}}=P_{k_{1}k_{2}}P_{k_{2}k_{3}}$ and $P_{\gamma _{k}\gamma
_{k}^{2}}P_{\gamma _{k}^{2}\gamma _{k}^{3}}=P_{k_{2}k_{3}}P_{k_{3}k_{4}}$,
whence the proper \emph{''deplanarizing operator lift''} of the term $%
P_{k\gamma _{k}}P_{\gamma _{k}\gamma _{k}^{2}}$ reads ($\zeta _{3}\in R$)
\begin{equation}
6P_{k\gamma _{k}}P_{\gamma _{k}\gamma _{k}^{2}}\rightarrow \zeta
_{3}P_{k_{1}k_{2}}P_{k_{2}k_{3}}+\left( 6-\zeta _{3}\right)
P_{k_{2}k_{3}}P_{k_{3}k_{4}}.
\end{equation}

Analogously, for the other terms of $H_{6,planar}$ (\ref{D6}) we obtain the
following proper \emph{''deplanarizing operator lifts''} ($\zeta _{4}\in R$%
):
\begin{equation}
\begin{array}{l}
6P_{\gamma _{k}\gamma _{k}^{2}}P_{\gamma _{k}^{2}\gamma _{k}^{3}}\rightarrow
\zeta _{4}P_{k_{2}k_{3}}P_{k_{3}k_{4}}+\left( 6-\zeta _{4}\right)
P_{k_{3}k_{4}}P_{k_{2}k_{3}}, \\
P_{k\gamma _{k}}P_{\gamma _{k}^{2}\gamma _{k}^{3}}\rightarrow
P_{k_{1}k_{2}}P_{k_{3}k_{4}}, \\
P_{k\gamma _{k}}P_{\gamma _{k}\gamma _{k}^{2}}P_{\gamma _{k}^{2}\gamma
_{k}^{3}}\rightarrow P_{k_{1}k_{2}}P_{k_{2}k_{3}}P_{k_{3}k_{4}}, \\
P_{\gamma _{k}^{2}\gamma _{k}^{3}}P_{\gamma _{k}\gamma _{k}^{2}}P_{k\gamma
_{k}}\rightarrow P_{k_{3}k_{4}}P_{k_{2}k_{3}}P_{k_{1}k_{2}}.
\end{array}
\end{equation}

Finally, considering $\frac{1}{N^{3}}\Sigma _{k_{1}k_{2}k_{3}k_{4}}\left(
\gamma \right) $ as the linking variable part, and therefore imposing the
symmetry of the spin part under the site index exchange
\begin{equation}
\left( k_{1},k_{2},k_{3},k_{4}\right) \leftrightarrow \left(
k_{4},k_{3},k_{2},k_{1}\right) ,
\end{equation}
we may write the following expression for the non-planar, $3$-loop $su(2)$
spin-chain Hamiltonian ($\alpha _{1},\alpha _{2}\in R$):
\begin{eqnarray}
H_{6}\left( \alpha _{1},\alpha _{2}\right) &=&\frac{4}{N^{3}}\sum_{\left(
k_{1},k_{2},k_{3},k_{4}\right) \in Z_{L}^{4}}  \notag \\
&&\left[
\begin{array}{l}
15+\alpha _{1}\left( P_{k_{1}k_{2}}+P_{k_{3}k_{4}}\right) -2\left( \alpha
_{1}+13\right) P_{k_{2}k_{3}}+ \\
+\alpha _{2}\left(
P_{k_{1}k_{2}}P_{k_{2}k_{3}}+P_{k_{3}k_{4}}P_{k_{2}k_{3}}\right) + \\
+\left( 6-\alpha _{2}\right) \left(
P_{k_{2}k_{3}}P_{k_{1}k_{2}}+P_{k_{2}k_{3}}P_{k_{3}k_{4}}\right) + \\
+P_{k_{1}k_{2}}P_{k_{3}k_{4}}-P_{k_{1}k_{2}}P_{k_{2}k_{3}}P_{k_{3}k_{4}}-P_{k_{3}k_{4}}P_{k_{2}k_{3}}P_{k_{1}k_{2}}
\end{array}
\right] \Sigma _{k_{1}k_{2}k_{3}k_{4}}\left( \gamma \right) .  \notag \\
&&  \label{60}
\end{eqnarray}
Notice that $H_{6}$ is completely symmetric under the site index exchange
\begin{equation}
\left( k_{1},k_{2},k_{3},k_{4}\right) \leftrightarrow \left(
k_{4},k_{3},k_{2},k_{1}\right) ,
\end{equation}
as it has to be from the site-index structure obtained from the linking part
$\Sigma _{k_{1}\gamma _{k_{2}}}\Sigma _{k_{2}\gamma _{k_{3}}}\Sigma
_{k_{3}\gamma _{k_{4}}}$ of the Hamiltonian.

It is now convenient to introduce a non-planar generalization of the
short-hand notation defined in Eq. (F.1) of \cite{Beisert:2003tq} and
previously reported in Eq. (\ref{notation-planar}), in the following way:
\begin{eqnarray}
&&
\begin{array}{l}
\left\{ \left( k_{1}k_{2},k_{2}k_{3},k_{3}k_{4},k_{1}k_{4},...\right)
_{P};\left( k_{1}k_{3},k_{1}k_{4},...\right) _{\gamma }\right\} \equiv \\
\equiv \sum_{k_{1},k_{2},k_{3},k_{4},...\in
Z_{L}}P_{k_{1}k_{2}}P_{k_{2}k_{3}}P_{k_{3}k_{4}}P_{k_{1}k_{4}}...\Sigma
_{k_{1}\gamma _{k_{3}}}\Sigma _{k_{1}\gamma _{k_{4}}}...,
\end{array}
\notag \\
&&
\end{eqnarray}
where we have the notational identification
\begin{equation}
\left\{ ;\left( k_{1}k_{3},k_{1}k_{4},...\right) _{\gamma }\right\} \equiv
\left\{ 1;\left( k_{1}k_{3},k_{1}k_{4},...\right) _{\gamma }\right\} \equiv
\sum_{k_{1},k_{3},k_{4},...\in Z_{L}}\Sigma _{k_{1}\gamma _{k_{3}}}\Sigma
_{k_{1}\gamma _{k_{4}}}...\text{ \ .}
\end{equation}
With such a notation the \emph{Ansatz} (\ref{60}) for the full non-planar, $%
3 $-loop $su(2)$ spin-chain Hamiltonian becomes
\begin{gather}
H_{6}\left( \alpha _{1},\alpha _{2}\right) =  \notag \\
\notag \\
=\frac{4}{N^{3}}\left[
\begin{array}{l}
15+\alpha _{1}\left( k_{1}k_{2}+k_{3}k_{4}\right) _{P}-2\left( \alpha
_{1}+13\right) \left( k_{2}k_{3}\right) _{P}+ \\
+\alpha _{2}\left[ \left( k_{1}k_{2},k_{2}k_{3}\right) _{P}+\left(
k_{3}k_{4},k_{2}k_{3}\right) _{P}\right] + \\
+\left( 6-\alpha _{2}\right) \left[ \left( k_{2}k_{3},k_{1}k_{2}\right)
_{P}+\left( k_{2}k_{3},k_{3}k_{4}\right) _{P}\right] + \\
+\left( k_{1}k_{2},k_{3}k_{4}\right) _{P}-\left(
k_{1}k_{2},k_{2}k_{3},k_{3}k_{4}\right) _{P}-\left(
k_{3}k_{4},k_{2}k_{3},k_{1}k_{2}\right) _{P}; \\
\left( k_{1}k_{2},k_{2}k_{3},k_{3}k_{4}\right) _{\gamma }
\end{array}
\right] .  \label{63}
\end{gather}

A completely analogous approach with the same steps can be performed for the
$4$-loop order, the only difference being that, already at the planar level,
expressed by Eq. (\ref{D8}), the Hamiltonian depends on an undetermined real
parameter $\beta $; the final result for the \emph{Ansatz }on the full
non-planar, $4$-loop $su(2)$ spin-chain Hamiltonian is the following%
\footnote{%
We use the semicolon in the arguments of $H_{8}$ to indicate the fact that
the real parameter $\beta $ survives also at the $4$-loop, planar level.} ($%
\eta _{1},\eta _{2},\eta _{3},\eta _{4},\beta \in R$):
\begin{gather}
H_{8}\left( \eta _{1},\eta _{2},\eta _{3},\eta _{4};\beta \right) =\frac{1}{%
N^{4}}\cdot   \notag  \label{64} \\
\notag \\
\cdot \left\{
\begin{array}{l}
-560+\left( 518+2\beta -\eta _{1}\right) \left( k_{1}k_{2}+k_{4}k_{5}\right)
_{P}+\eta _{1}\left( k_{2}k_{3}+k_{3}k_{4}\right) _{P}+ \\
+\eta _{2}\left[ \left( k_{1}k_{2},k_{2}k_{3}\right) _{P}+\left(
k_{4}k_{5},k_{3}k_{4}\right) _{P}\right] +\eta _{3}\left[ \left(
k_{2}k_{3},k_{3}k_{4}\right) _{P}+\left( k_{3}k_{4},k_{2}k_{3}\right) _{P}%
\right] + \\
-\left( 266+4\beta +\eta _{2}+\eta _{3}\right) \left[ \left(
k_{3}k_{4},k_{4}k_{5}\right) _{P}+\left( k_{2}k_{3},k_{1}k_{2}\right) _{P}%
\right] + \\
-\left( 33+\beta \right) \left[ \left( k_{1}k_{2},k_{3}k_{4}\right)
_{P}+\left( k_{2}k_{3},k_{4}k_{5}\right) _{P}\right] -4\left(
k_{1}k_{2},k_{4}k_{5}\right) _{P}+ \\
+4\left[
\begin{array}{l}
\left( k_{1}k_{2},k_{2}k_{3},k_{4}k_{5}\right) _{P}+\left(
k_{1}k_{2},k_{3}k_{4},k_{4}k_{5}\right) _{P}+ \\
+\left( k_{1}k_{2},k_{4}k_{5},k_{3}k_{4}\right) _{P}+\left(
k_{2}k_{3},k_{1}k_{2},k_{4}k_{5}\right) _{P}
\end{array}
\right] + \\
+\eta _{4}\left[ \left( k_{1}k_{2},k_{2}k_{3},k_{3}k_{4}\right) _{P}+\left(
k_{4}k_{5},k_{3}k_{4},k_{2}k_{3}\right) _{P}\right] + \\
+\left( 78+2\beta -\eta _{4}\right) \left[ \left(
k_{2}k_{3},k_{3}k_{4},k_{4}k_{5}\right) _{P}+\left(
k_{3}k_{4},k_{2}k_{3},k_{1}k_{2}\right) _{P}\right] + \\
+\left( \beta -9\right) \left[
\begin{array}{l}
\left( k_{1}k_{2},k_{3}k_{4},k_{2}k_{3}\right) _{P}+\left(
k_{2}k_{3},k_{4}k_{5},k_{3}k_{4}\right) _{P}+ \\
+\left( k_{2}k_{3},k_{1}k_{2},k_{3}k_{4}\right) _{P}+\left(
k_{3}k_{4},k_{2}k_{3},k_{4}k_{5}\right) _{P}
\end{array}
\right] + \\
+\left( 1-\beta \right) \left[
\begin{array}{l}
\left( k_{1}k_{2},k_{2}k_{3},k_{4}k_{5},k_{3}k_{4}\right) _{P}+\left(
k_{1}k_{2},k_{4}k_{5},k_{3}k_{4},k_{2}k_{3}\right) _{P}+ \\
+\left( k_{2}k_{3},k_{1}k_{2},k_{3}k_{4},k_{4}k_{5}\right) _{P}+\left(
k_{3}k_{4},k_{2}k_{3},k_{1}k_{2},k_{4}k_{5}\right) _{P}
\end{array}
\right] + \\
+\left( 3-\beta \right) \left[ \left(
k_{2}k_{3},k_{1}k_{2},k_{3}k_{4},k_{2}k_{3}\right) _{P}+\left(
k_{3}k_{4},k_{2}k_{3},k_{4}k_{5},k_{3}k_{4}\right) _{P}\right] + \\
+2\beta \left[ \left( k_{1}k_{2},k_{3}k_{4},k_{2}k_{3},k_{4}k_{5}\right)
_{P}+\left( k_{2}k_{3},k_{1}k_{2},k_{4}k_{5},k_{3}k_{4}\right) _{P}\right] +
\\
-10\left[ \left( k_{1}k_{2},k_{2}k_{3},k_{3}k_{4},k_{4}k_{5}\right)
_{P}+\left( k_{4}k_{5},k_{3}k_{4},k_{2}k_{3},k_{1}k_{2}\right) _{P}\right] ;
\\
\left( k_{1}k_{2},k_{2}k_{3},k_{3}k_{4},k_{4}k_{5}\right) _{\gamma }
\end{array}
\right\} .  \notag \\
\label{644}
\end{gather}
Notice that $H_{8}$ is completely symmetric under the site index exchange
\begin{equation}
\left( k_{1},k_{2},k_{3},k_{4},k_{5}\right) \leftrightarrow \left(
k_{5},k_{4},k_{3},k_{2},k_{1}\right) ,
\end{equation}
as it has to be from the site-index structure obtained from the linking part
$\Sigma _{k_{1}\gamma _{k_{2}}}\Sigma _{k_{2}\gamma _{k_{3}}}\Sigma
_{k_{3}\gamma _{k_{4}}}\Sigma _{k_{4}\gamma _{k_{5}}}$ of the Hamiltonian.

Whence, trivially noticing that the above considered \emph{%
''deplanarization''} procedure \ has no effects on the tree level, we may
say that Eqs. (\ref{34}), (\ref{36}), (\ref{63}) and (\ref{644}) are a
consistent full non-planar generalization of the corresponding planar
formulae, respectively given by Eqs. (\ref{D2}), (\ref{D4}), (\ref{D6}) and (%
\ref{D8}).

However, while Eqs. (\ref{34}) and (\ref{36}) (respectively corresponding to
the full non-planar $1$- and $2$-loop level) are exactly determined, notice
that Eqs. (\ref{63}) and (\ref{644}) (respectively corresponding to the full
non-planar $3$- and $4$-loop level) contain undetermined free real
parameters. Therefore are actually not completely fixed. Such free real
parameters (surviving, in the $4$-loop case, also at the planar level)
should be fixed by matching with some known results about the non-planar, $3$%
- and $4$-loop level for the $su(2)$-symmetric operatorial sector of the
considered 4-d. $\mathcal{N}=4$ SYM theory; unfortunately, at the moment
such (independently obtained) full non-planar higher-loop results are
unavailable.

By the way, in order to completely fix the free parameters introduced by the
general \emph{deplanarization procedure}, we may formulate an additional
assumption, that we are going to call \emph{hypothesis of ''symmetrization
of deplanarizing operator splittings''}. This conjecture has to be applied
\emph{after} the symmetrization of the non-planar terms with respect to the
peculiar renaming of spin-chain site indices, which is given by (\ref
{exchange}) and determined by the linking part of the Hamiltonian expressed
by (\ref{38bis}); it amounts to say that \emph{each} of the sets of
non-planar terms arising from a considered planar term in the
deplanarization procedure will \emph{equally} contribute in the planar limit
$N\rightarrow \infty $. For example, if, after the symmetrization with
respect to (\ref{exchange}), a planar term $\Im $ is deplanarized by the $3$%
-fold splitting
\begin{equation}
\Im \rightarrow a_{1}\Im _{M_{1}}+a_{2}\Im _{M_{2}}+a_{3}\Im _{M_{3}},
\label{split1}
\end{equation}
where $\Im _{M_{1}}$, $\Im _{M_{2}}$ and $\Im _{M_{3}}$ are sets consisting
of $M_{1}$, $M_{2}$ and $M_{3}$ non-planar terms made by permutation
operators, then we will assume that
\begin{equation}
M_{1}a_{1}=M_{2}a_{2}=M_{3}a_{3}.
\end{equation}
Hence the contribution of $\Im _{M_{1}}$, $\Im _{M_{2}}$ and $\Im _{M_{3}}$
to the planar limit $\Im $ is the \emph{same}, and therefore the operator
splitting given by (\ref{split1}) may be considered \emph{symmetric}.

By applying such a simple additional conjecture to the $3$- and $4$-loop
orders, we obtain that the \emph{Ans\"{a}tze} (\ref{63}) and (\ref{644}) for
the non-planar, $3$- and $4$-loop $su(2)$ spin-chain Hamiltonians become
\emph{completely determined}, respectively reading as follows:
\begin{gather}
H_{6,non-planar}=  \notag \\
=\frac{4}{N^{3}}\left[
\begin{array}{l}
15-\frac{13}{2}\left( k_{1}k_{2}+k_{3}k_{4}\right) _{P}-13\left(
k_{2}k_{3}\right) _{P}+ \\
+3\left[ \left( k_{1}k_{2},k_{2}k_{3}\right) _{P}+\left(
k_{3}k_{4},k_{2}k_{3}\right) _{P}+\right. \\
\left. +\left( k_{2}k_{3},k_{1}k_{2}\right) _{P}+\left(
k_{2}k_{3},k_{3}k_{4}\right) _{P}\right] + \\
+\left( k_{1}k_{2},k_{3}k_{4}\right) _{P}-\left(
k_{1}k_{2},k_{2}k_{3},k_{3}k_{4}\right) _{P}-\left(
k_{3}k_{4},k_{2}k_{3},k_{1}k_{2}\right) _{P}; \\
\left( k_{1}k_{2},k_{2}k_{3},k_{3}k_{4}\right) _{\gamma }
\end{array}
\right] ;  \label{3-loop-fixed}
\end{gather}
\begin{gather}
H_{8,non-planar}\left( \beta \right) =  \notag \\
\notag \\
=\frac{1}{N^{4}}\left\{
\begin{array}{l}
-560+\left( 259+\beta \right) \left(
k_{1}k_{2}+k_{4}k_{5}+k_{2}k_{3}+k_{3}k_{4}\right) _{P}+ \\
-\left( \frac{266}{3}+\frac{4}{3}\beta \right) \left[
\begin{array}{l}
\left( k_{1}k_{2},k_{2}k_{3}\right) _{P}+\left( k_{4}k_{5},k_{3}k_{4}\right)
_{P}+ \\
+\left( k_{2}k_{3},k_{3}k_{4}\right) _{P}+\left(
k_{3}k_{4},k_{2}k_{3}\right) _{P}+ \\
+\left( k_{3}k_{4},k_{4}k_{5}\right) _{P}+\left(
k_{2}k_{3},k_{1}k_{2}\right) _{P}
\end{array}
\right] + \\
-\left( 33+\beta \right) \left[ \left( k_{1}k_{2},k_{3}k_{4}\right)
_{P}+\left( k_{2}k_{3},k_{4}k_{5}\right) _{P}\right] -4\left(
k_{1}k_{2},k_{4}k_{5}\right) _{P}+ \\
+4\left[
\begin{array}{l}
\left( k_{1}k_{2},k_{2}k_{3},k_{4}k_{5}\right) _{P}+\left(
k_{1}k_{2},k_{3}k_{4},k_{4}k_{5}\right) _{P}+ \\
+\left( k_{1}k_{2},k_{4}k_{5},k_{3}k_{4}\right) _{P}+\left(
k_{2}k_{3},k_{1}k_{2},k_{4}k_{5}\right) _{P}
\end{array}
\right] + \\
+\left( 39+\beta \right) \left[
\begin{array}{l}
\left( k_{1}k_{2},k_{2}k_{3},k_{3}k_{4}\right) _{P}+\left(
k_{4}k_{5},k_{3}k_{4},k_{2}k_{3}\right) _{P}+ \\
+\left( k_{2}k_{3},k_{3}k_{4},k_{4}k_{5}\right) _{P}+\left(
k_{3}k_{4},k_{2}k_{3},k_{1}k_{2}\right) _{P}
\end{array}
\right] + \\
+\left( \beta -9\right) \left[
\begin{array}{l}
\left( k_{1}k_{2},k_{3}k_{4},k_{2}k_{3}\right) _{P}+\left(
k_{2}k_{3},k_{4}k_{5},k_{3}k_{4}\right) _{P}+ \\
+\left( k_{2}k_{3},k_{1}k_{2},k_{3}k_{4}\right) _{P}+\left(
k_{3}k_{4},k_{2}k_{3},k_{4}k_{5}\right) _{P}
\end{array}
\right] + \\
+\left( 1-\beta \right) \left[
\begin{array}{l}
\left( k_{1}k_{2},k_{2}k_{3},k_{4}k_{5},k_{3}k_{4}\right) _{P}+\left(
k_{1}k_{2},k_{4}k_{5},k_{3}k_{4},k_{2}k_{3}\right) _{P}+ \\
+\left( k_{2}k_{3},k_{1}k_{2},k_{3}k_{4},k_{4}k_{5}\right) _{P}+\left(
k_{3}k_{4},k_{2}k_{3},k_{1}k_{2},k_{4}k_{5}\right) _{P}
\end{array}
\right] + \\
+\left( 3-\beta \right) \left[ \left(
k_{2}k_{3},k_{1}k_{2},k_{3}k_{4},k_{2}k_{3}\right) _{P}+\left(
k_{3}k_{4},k_{2}k_{3},k_{4}k_{5},k_{3}k_{4}\right) _{P}\right] + \\
+2\beta \left[ \left( k_{1}k_{2},k_{3}k_{4},k_{2}k_{3},k_{4}k_{5}\right)
_{P}+\left( k_{2}k_{3},k_{1}k_{2},k_{4}k_{5},k_{3}k_{4}\right) _{P}\right] +
\\
-10\left[ \left( k_{1}k_{2},k_{2}k_{3},k_{3}k_{4},k_{4}k_{5}\right)
_{P}+\left( k_{4}k_{5},k_{3}k_{4},k_{2}k_{3},k_{1}k_{2}\right) _{P}\right] ;
\\
\left( k_{1}k_{2},k_{2}k_{3},k_{3}k_{4},k_{4}k_{5}\right) _{\gamma }
\end{array}
\right\} .  \label{4-loop-fixed}
\end{gather}

\section{Conclusions and perspectives}

In summary, we worked out a general procedure of full \emph{%
''deplanarization''} of planar results about $su(2)$ closed spin-chain
Hamiltonian, describing the dynamics of the $su(2)$ sector of the 4-d. $%
\mathcal{N}=4$ SYM theory. Such a method is based on proper \emph{%
''deplanarizing lifts''} of \ (products of) site permutation operators $P$'s
of the kind
\begin{equation}
\zeta P_{k,k+r}\longrightarrow \sum_{i=1}^{n\text{ (loop order)}}\zeta
_{i}P_{k_{i},k_{i}+r}\text{, such that }\sum_{i=1}^{n\text{ (loop order)}%
}\zeta _{i}=\zeta ,\text{ \ }r\in Z_{L},\zeta \in R,
\end{equation}
and in general
\begin{gather}
\zeta P_{k,k+r_{1}}P_{k,k+r_{2}}...\longrightarrow \sum_{i=1}^{n\text{ (loop
order)}}\zeta _{i}P_{k_{i},k_{i}+r_{1}}P_{k_{i},k_{i}+r_{2}}...\text{,}
\notag \\
\text{such that }\sum_{i=1}^{n\text{ (loop order)}}\zeta _{i}=\zeta ,\text{
\ }r_{1},r_{2},...\in Z_{L},\zeta _{i}\in R.
\end{gather}
The number of free parameters $\left\{ \zeta _{i}\right\} $ may be decreased
by observing that the assumed \emph{Ansatz} on the linking part of the
higher-loop, non-planar Hamiltonian determines a simmetry under a certain
exchange of site indices, expressed by Eq. (\ref{exchange}) and implied by $%
\Sigma _{k_{0}k_{1}...k_{n}}\left( \gamma \right) $ defined in (\ref{38bis}).

Furthermore, performing an (extremely reasonable) additional conjecture of \
\emph{''symmetrization of deplanarizing operator splittings''}, we were able
to completely fix \emph{all} free parameters introduced by the proposed
\emph{''deplanarization procedure''}. Thus, we wrote down some \emph{%
completely determined} (up to the free planar parameter $\beta $ at $4$-loop
order) expressions for the full non-planar, $3$- and $4$-loop $su(2)$
spin-chain Hamiltonians, respectively given by Eqs. (\ref{3-loop-fixed}) and
(\ref{4-loop-fixed}).

Such a ''deplanarizing'' procedure is perfectly compatible with
(independently obtained) known results at the $1$- and $2$-loop, non-planar
level \cite{Bellucci:2004ru,Bellucci:2004qx,Bellucci:2005ma1}; moreover, by
construction, the above-formulated \emph{Ans\"{a}tze} are \emph{planarly
consistent}, i.e. they have the correct planar limit, matching the known
results reported in the literature (see e.g. \cite
{Beisert:2003tq,Beisert:2003hl}).

It is worth noticing that, imposing only invariance under (\ref{exchange}),
the general $3$- and $4$-loop \emph{Ans\"{a}tze} (\ref{63}) and (\ref{644})
do contain \emph{free} parameters, easily fixable by matching procedures
with (independently obtained) non-planar higher-loop results, achieved by
using Feynman diagrams approach on the gauge theory side, or performing
perturbative string calculations. Unfortunately, at the moment such results
to match with are unavailable, but nonetheless the above-proposed \emph{%
Ans\"{a}tze} express the most general form of non-planar Hamiltonians. Also,
notice that such free undetermined parameters should \emph{not} affect the
properties and structure of relevant physical quantities, such as the
spectrum; as previously mentioned, this is claimed in the known literature
(see e.g. \cite{Beisert:2004di}) to hold at the planar level and, due to the
the planar consistence, it seems natural and reasonable to conjecturally
extend such a claim also at the non-planar level.

Nonetheless, due to the failure of the elegant \emph{''spin
edge-differences''} \emph{Ans\"{a}tze} (\ref{41}) for the explicit form of
the $n$-loop, non-planar $su(2)$ spin-chain Hamiltonian, the \emph{%
''deplanarizing''} method here proposed seems, at the moment, the only one
capable of guaranteeing planar consistency and, in principle, full
testability for the higher-loop, non-planar \emph{Ans\"{a}tze}\footnote{%
Notice also that the proposed \emph{Ans\"{a}tze} for the $3$- and $4$-loop,
non-planar Hamiltonians can always be cast in terms of the two-site planar
one-loop Hamiltonian $H_{kl}$. This fact allows us to claim that such \emph{%
Ans\"{a}tze} do formally hold at least also for the $su(3)$ sector of the
4-d. $\mathcal{N}=4$ SYM theory. Similarly to the $su(2)$ one, such a sector
is made out by local gauge invariant SYM operators which are generated by
three holomorphic (multi)trace operators built out of two complex SYM
scalars $X=\phi _{5}+\mathrm{i}\phi _{6}$, $Y=\phi _{3}+\mathrm{i}\phi _{4}$
and $Z=\phi _{1}+\mathrm{i}\phi _{2}$, of typical form
\begin{equation*}
\mathcal{O}=\mathop{\mathrm{Tr}}\nolimits(XZYYZ\dots )\mathop{\mathrm{Tr}}%
\nolimits(XYXZ\dots )\mathop{\mathrm{Tr}}\nolimits(\dots )\dots
\end{equation*}
}.

Also notice that, by construction, \emph{Ans\"{a}tze} (\ref{63}) and (\ref
{644}) show an explicit \emph{full factorization} in the spin and
chain-splitting parts; as already pointed out in \cite{Bellucci:2005ma1},
such a property is expected to hold at \emph{every} loop order, since the
Hilbert space of the spin bit model is always given by the direct product
(modulo the action of the permutation group $S_{L}$) of the spin space and
the linking variable $\gamma $-space.

As previously considered, attention must also be paid to the fact that,
while (both at non-planar and planar levels) the $1$- and $2$-loop formulae
for the $su(2)$ spin-chain Hamiltonian are \emph{linear} in the site
permutation operators $P$'s, the $3$- and $4$-loop level expressions, both
in the planar case (see (\ref{D6}) and (\ref{D8})) and in the non-planar
case (see Eqs. (\ref{63}) and (\ref{644})), do seem to show a \emph{%
non-linearity} (and \emph{non-linearizability}) in $P$'s. For example, the
\emph{non-linearizability} of the planar, $3$-loop $su(2)$ spin-chain
Hamiltonian (\ref{D6}) caused the failure of the elegant and geometrically
meaningful \emph{''spin edge-differences''} approach to higher-loop \emph{%
Ans\"{a}tze}.

Thus, the \emph{non-linearity} (and \emph{non-linearizability}) in site
permutation operators seems to be a crucial and fundamental feature,
starting to hold at the $3$-loop order, of the spin part of the $su(2)$
spin-chain Hamiltonian for the spin bit model, underlying the dynamics of
the $su(2)$ sector of the 4-d. $\mathcal{N}=4$ SYM theory. Such a \emph{%
''breakdown'' of ''permutational linearizability''} at the $3$-loop level
would give rise, by means of the AdS/CFT correspondence \cite
{Maldacena:1998re,Gubser:1998bc}, to some ''new'' features in the dynamics
of the (closed) superstrings in the bulk of $AdS_{5}\times S^{5}$ (for the
first evidences from $3$-loop calculations, see e.g. \cite
{Beisert:2003tq,Beisert:2003hl}; for further subsequent developments see
e.g. \cite{Tseytlin:2004xa,Tseytlin:2003ac,Ryzhov:2004-1,Ryzhov:2004-2}%
).\smallskip

Finally, we notice that it would be interesting, following recent research
directions, to extend such a \emph{''deplanarization''} method to other
operatorial sectors of the 4-d. $\mathcal{N}=4$ SYM theory \cite
{Bellucci:2005np1,Bellucci:2005sm1}; unfortunately, even restricting the
consideration to the planar level, only $su(2)$ anomalous dimension
operators are presently known beyond $1$-loop.

\subsection*{Acknowledgements}

We benefited from useful discussions with P.-Y. Casteill and C. Sochichiu.
Special thanks are extended to F. Morales, for careful reading of the
manuscript and stimulating suggestions. This research was partially
supported by the European Community's Marie Curie Research Training Network
under contract MRTN-CT-2004-005104 Forces Universe, as well as by
INTAS-00-00254 grant.
\begin{equation*}
\end{equation*}

\subsection*{Appendix I}

\subsection*{$su(2)$\ sector: applying permutational identities to
ultra-localization by Pauli matrices}

In the $su(2)$-symmetry operatorial sector of the 4-d. $\mathcal{N}=4$ SYM
theory the spin operators form a spin $s=\frac{1}{2}$ (representation of
the) $su(2)$ algebra on each site of the spin-chain\footnote{%
As usual, we will respectively denote with $\left[ \cdot ,\cdot \right] $
and $\left\{ \cdot ,\cdot \right\} $ the commutator and anticommutator of
(matrix representations of) operators.}
\begin{equation}
\left[ \widehat{S}^{i},\widehat{S}^{j}\right] =i\epsilon ^{ijk}\widehat{S}%
^{k},
\end{equation}
where here and in the following supescript indices range in $\left\{
1,2,3\right\} $ (3-d. spatial spin degrees of freedom), and eigenvalues of
spin operators $\widehat{S}^{i}$ (denoted with $S^{i}$) take values in $%
Z_{2} $, i.e. in the integer numeric field with period $2$.

Imposing that such an $su(2)$ symmetry be an \emph{ultra-local} one, i.e.
that spin operators belonging to different (also nearest-neighbouring) sites
of the spin-chain always commute, it is then possible to say that in the
case in which (subscript denotes the site position)
\begin{equation}
S_{k}^{i}\in Z_{2}\ \forall \left( k,i\right) \in Z_{L}\times \left\{
1,2,3\right\} ,
\end{equation}
we can irreducibly represent the ultra-localized $s=\frac{1}{2}$ $su(2)$
algebra
\begin{equation}
\left[ \widehat{S}_{k}^{i},\widehat{S}_{k_{2}}^{j}\right] =i\delta
_{kl}\epsilon ^{ijm}\widehat{S}_{k}^{m}  \label{15}
\end{equation}
($\delta _{kl}$ denoting the usual Kr\"{o}necker delta) by the usual Pauli $%
\sigma $-matrices, defined as
\begin{equation}
\widehat{S}_{k}^{i}=:\frac{1}{2}\sigma _{k}^{i}.  \label{16}
\end{equation}
Therefore, Eq. (\ref{15}) has the irreducible matrix representation
\begin{equation}
\left[ \sigma _{k}^{i},\sigma _{l}^{j}\right] =2i\delta _{kl}\epsilon
^{ijm}\sigma _{k}^{m},  \label{17}
\end{equation}
expressing the ultra-localization of the Pauli $\sigma $-matrix algebra
(i.e. the irreducible representation of the ultra-localized $s=\frac{1}{2}$ $%
su(2)$ algebra) on the spin-chain sites.

Using Eqs. (\ref{15})-(\ref{17}), the action of the permutation operator $%
P_{kl}=P_{\widehat{S}_{k}\widehat{S}_{l}}$ may be represented in the
following way
\begin{equation}
P_{kl}=\frac{1}{2}\left( 1+\overrightarrow{\sigma _{k}}\cdot \overrightarrow{%
\sigma _{l}}\right) ,\text{ \ \ \ \ \ }\left( k,l\right) \in \left\{
1,...,L\right\} ^{2},k\neq l,  \label{18}
\end{equation}
with $\overrightarrow{\sigma _{k}}$ being the matrix 3-vector $\left( \sigma
_{k}^{1},\sigma _{k}^{2},\sigma _{k}^{3}\right) $ of Pauli $\sigma $%
-matrices on the $k$-th spin-chain site, and $\cdot $ denoting the usual
Euclidean scalar product between such matrix 3-vectors (based on the
standard ''row-columns'' matrix product, denoted in the following with $%
\times $).

At this point, using Eq. (\ref{18}), we may reformulate the permutational
identities obtained in Section 2, specializing them to the case of the \emph{%
ultra-localized} $su(2)$-symmetric operatorial sector of the 4-d. $\mathcal{N%
}=4$ SYM theory, with such a symmetry irreducibly represented by Pauli $%
\sigma $-matrices.

Starting from the identity (\ref{1st-id}), we thence get ($k_{1}\neq
k_{2}\neq k_{3}$ throughout)
\begin{equation}
\begin{array}{l}
\overrightarrow{\sigma _{k_{1}}}\cdot \overrightarrow{\sigma _{k_{2}}}+%
\overrightarrow{\sigma _{k_{1}}}\cdot \overrightarrow{\sigma _{k_{3}}}%
+\left( \overrightarrow{\sigma _{k_{1}}}\cdot \overrightarrow{\sigma _{k_{2}}%
}\right) \left( \overrightarrow{\sigma _{k_{1}}}\cdot \overrightarrow{\sigma
_{k_{3}}}\right) = \\
=\overrightarrow{\sigma _{k_{2}}}\cdot \overrightarrow{\sigma _{k_{3}}}+%
\overrightarrow{\sigma _{k_{1}}}\cdot \overrightarrow{\sigma _{k_{2}}}%
+\left( \overrightarrow{\sigma _{k_{2}}}\cdot \overrightarrow{\sigma _{k_{3}}%
}\right) \left( \overrightarrow{\sigma _{k_{1}}}\cdot \overrightarrow{\sigma
_{k_{2}}}\right) = \\
=\overrightarrow{\sigma _{k_{1}}}\cdot \overrightarrow{\sigma _{k_{3}}}+%
\overrightarrow{\sigma _{k_{2}}}\cdot \overrightarrow{\sigma _{k_{3}}}%
+\left( \overrightarrow{\sigma _{k_{1}}}\cdot \overrightarrow{\sigma _{k_{3}}%
}\right) \left( \overrightarrow{\sigma _{k_{2}}}\cdot \overrightarrow{\sigma
_{k_{3}}}\right) ,
\end{array}
\end{equation}
implying
\begin{equation}
\overrightarrow{\sigma _{k_{1}}}\cdot \overrightarrow{\sigma _{k_{3}}}-%
\overrightarrow{\sigma _{k_{2}}}\cdot \overrightarrow{\sigma _{k_{3}}}%
=\left( \overrightarrow{\sigma _{k_{2}}}\cdot \overrightarrow{\sigma _{k_{3}}%
}\right) \left( \overrightarrow{\sigma _{k_{1}}}\cdot \overrightarrow{\sigma
_{k_{2}}}\right) -\left( \overrightarrow{\sigma _{k_{1}}}\cdot
\overrightarrow{\sigma _{k_{2}}}\right) \left( \overrightarrow{\sigma
_{k_{1}}}\cdot \overrightarrow{\sigma _{k_{3}}}\right) .  \label{20}
\end{equation}
Since Eq. (\ref{17}) implies ($\wedge $ denotes the vector product between
matrix 3-vectors)
\begin{equation}
\left[ \left( \overrightarrow{\sigma _{k_{1}}}\cdot \overrightarrow{\sigma
_{k_{2}}}\right) ,\left( \overrightarrow{\sigma _{k_{2}}}\cdot
\overrightarrow{\sigma _{k_{3}}}\right) \right] =2i\left( \overrightarrow{%
\sigma _{k_{1}}}\wedge \overrightarrow{\sigma _{k_{3}}}\right) \cdot
\overrightarrow{\sigma _{k_{2}}},  \label{21}
\end{equation}
Eq. (\ref{20}) may be rewritten as
\begin{equation}
\begin{array}{l}
\overrightarrow{\sigma _{k_{1}}}\cdot \overrightarrow{\sigma _{k_{3}}}\left(
1-\overrightarrow{\sigma _{k_{1}}}\cdot \overrightarrow{\sigma _{k_{2}}}%
\right) -\left( 1-\overrightarrow{\sigma _{k_{1}}}\cdot \overrightarrow{%
\sigma _{k_{2}}}\right) \overrightarrow{\sigma _{k_{2}}}\cdot
\overrightarrow{\sigma _{k_{3}}}= \\
=2i\left[ \left( \overrightarrow{\sigma _{k_{1}}}\wedge \overrightarrow{%
\sigma _{k_{3}}}\right) \cdot \overrightarrow{\sigma _{k_{2}}}+\left(
\overrightarrow{\sigma _{k_{2}}}\wedge \overrightarrow{\sigma _{k_{3}}}%
\right) \cdot \overrightarrow{\sigma _{k_{1}}}\right] .
\end{array}
\end{equation}

Furthermore, considering the permutational identity (\ref{2nd-id}), we
obtain ($k_{1}\neq k_{2}\neq k_{3}$ throughout)
\begin{equation}
\overrightarrow{\sigma _{k_{1}}}\cdot \overrightarrow{\sigma _{k_{3}}}+\frac{%
1}{2}\left\{ \left( \overrightarrow{\sigma _{k_{1}}}\cdot \overrightarrow{%
\sigma _{k_{2}}}\right) ,\left( \overrightarrow{\sigma _{k_{2}}}\cdot
\overrightarrow{\sigma _{k_{3}}}\right) \right\} =0,
\end{equation}
which, by means of Eq. (\ref{21}), may be rewritten in the following way
\begin{equation}
\overrightarrow{\sigma _{k_{1}}}\cdot \overrightarrow{\sigma _{k_{3}}}%
+\left( \overrightarrow{\sigma _{k_{1}}}\cdot \overrightarrow{\sigma _{k_{2}}%
}\right) \left( \overrightarrow{\sigma _{k_{2}}}\cdot \overrightarrow{\sigma
_{k_{3}}}\right) -i\left( \overrightarrow{\sigma _{k_{1}}}\wedge
\overrightarrow{\sigma _{k_{3}}}\right) \cdot \overrightarrow{\sigma _{k_{2}}%
}=0.  \label{24}
\end{equation}

Finally, using Eqs. (\ref{18}) and (\ref{21}), the permutational identity (%
\ref{3rd-id}) may be put in the form ($k_{1}\neq k_{2}\neq k_{3}\neq k_{4}$
throughout)
\begin{equation}
\left( 1-\overrightarrow{\sigma _{k_{1}}}\cdot \overrightarrow{\sigma
_{k_{2}}}\right) \left( \overrightarrow{\sigma _{k_{1}}}\cdot
\overrightarrow{\sigma _{k_{3}}}+\overrightarrow{\sigma _{k_{3}}}\cdot
\overrightarrow{\sigma _{k_{2}}}\right) \left( 1+\overrightarrow{\sigma
_{k_{4}}}\cdot \overrightarrow{\sigma _{k_{1}}}+\overrightarrow{\sigma
_{k_{4}}}\cdot \overrightarrow{\sigma _{k_{2}}}+\overrightarrow{\sigma
_{k_{4}}}\cdot \overrightarrow{\sigma _{k_{3}}}\right) =0,
\end{equation}
or, equivalently, by the use of Eq. (\ref{24}):
\begin{equation}
\begin{array}{l}
\left[ 2\overrightarrow{\sigma _{k_{2}}}\cdot \left( \overrightarrow{\sigma
_{k_{3}}}+\overrightarrow{\sigma _{k_{4}}}\right) +i\left( \overrightarrow{%
\sigma _{k_{2}}}\wedge \overrightarrow{\sigma _{k_{3}}}\right) \cdot
\overrightarrow{\sigma _{k_{4}}}+i\left( \overrightarrow{\sigma _{k_{2}}}%
\wedge \overrightarrow{\sigma _{k_{4}}}\right) \cdot \overrightarrow{\sigma
_{k_{3}}}\right] \times \\
\times \left( 1+\overrightarrow{\sigma _{k_{1}}}\cdot \overrightarrow{\sigma
_{k_{2}}}+\overrightarrow{\sigma _{k_{1}}}\cdot \overrightarrow{\sigma
_{k_{3}}}+\overrightarrow{\sigma _{k_{1}}}\cdot \overrightarrow{\sigma
_{k_{4}}}\right) =0.
\end{array}
\end{equation}

In general, we may obtain many other equations involving Pauli $\sigma $%
-matrices ultra-localized on the spin-chain sites, by considering the
general form (\ref{gen-struct}) of the permutational identities previously
obtained, and using Eq. (\ref{18}):
\begin{equation}
\left( \sum_{\pi \in S_{M}}\left( -\right) ^{\sigma _{\pi }}P_{\pi }\right)
_{\text{with }P_{kl}=\frac{1}{2}\left( 1+\overrightarrow{\sigma _{k}}\cdot
\overrightarrow{\sigma _{l}}\right) ,k\neq l}=0.
\end{equation}
It is therefore possible to argue that such a ''tower'' of (reciprocally
independent) matrix equations, becoming more and more involved with the
increasing of $k_{3}$, represents a sort of \emph{''generalized Fierz
identities''} for Pauli $\sigma $-matrices, related to the spin $s=\frac{1}{2%
}$ irreducible representation of the \emph{ultra-localized} $su(2)$ algebra.
\begin{equation*}
\end{equation*}

\subsection*{Appendix II}

\subsection*{The planar limit of permutational identities}

Always focussing on the $su(2)$ sector of 4-d. $\mathcal{N}=4$ SYM theory,
the result (\ref{gen-struct}) may be considered as a full non-planar
generalization of previously known planar permutational identities, obtained
in \cite{Beisert:2003tq}.

To explicitly show this, let us report Eq. (F.2) of \cite{Beisert:2003tq};
in the notation specified by (\ref{notation-planar}) it is
\begin{equation}
\begin{array}{l}
\left\{ ...,n,n\pm 1,n,...\right\} + \\
-\left\{ ...,...\right\} +\left\{ ...,n,...\right\} +\left\{ ...,n\pm
1,...\right\} + \\
-\left\{ ...,n,n\pm 1,...\right\} -\left\{ ...,n\pm 1,n,...\right\} =0,\text{
\ }n\in Z.
\end{array}
\label{29}
\end{equation}
Therefore, it may be explicitly checked that
\begin{eqnarray}
&&
\begin{array}{l}
\left\{ 0,\pm 1,0\right\} -\left\{ {}\right\} +\left\{ 0\right\} +\left\{
\pm 1\right\} -\left\{ 0,\pm 1\right\} -\left\{ \pm 1,0\right\}
=0\Leftrightarrow \\
\Leftrightarrow \sum_{k\in Z_{L}}\left[
1-P_{k,k+1}-P_{k,k+2}-P_{k+1,k+2}+P_{k,k+1}P_{k+1,k+2}+P_{k+1,k+2}P_{k,k+1}%
\right] =0,
\end{array}
\notag \\
&&  \label{30}
\end{eqnarray}
which is nothing but the canonical planar limit of Eq.(\ref{2nd-id}), summed
on all $k\in Z_{L}$. This result is perfectly consistent, because, as it is
claimed in \cite{Beisert:2003tq}, the permutational identity (\ref{29}) is
\emph{''due to the impossibility of antisymmetrizing three sites in }$SU(2)$%
\emph{''}. Whence Eq. (\ref{30}) simply implies that Eq. (\ref{2nd-id}) is
nothing but the non-canonical, non-planar generalization of Eq. (\ref{29}).

Furthermore, using periodic boundary conditions on spin-chain, and
eventually applying (iterated) \emph{''pull-back''} or \emph{''push-forward''%
} of site indices by a permutation $\gamma \in S_{L}$, it may be shown that
Eq. (\ref{29}) actually does not depend on $n\in Z$ and on the choice of ''$%
\pm $'' at all, and it is always possible (without loss of generality) to
consider $n=0$, and choose ''$+$''.

Thus, Eq. (\ref{30}) may be rewritten as ($\forall n\in Z$)
\begin{eqnarray}
&&
\begin{array}{l}
\left\{ n,n\pm 1,n\right\} -\left\{ {}\right\} +\left\{ n\right\} +\left\{
n\pm 1\right\} -\left\{ n,n\pm 1\right\} -\left\{ n\pm 1,n\right\}
=0\Leftrightarrow \\
\Leftrightarrow \sum_{k\in Z_{L}}\left[
1-P_{k,k+1}-P_{k,k+2}-P_{k+1,k+2}+P_{k,k+1}P_{k+1,k+2}+P_{k+1,k+2}P_{k,k+1}%
\right] =0.
\end{array}
\notag \\
&&  \label{31}
\end{eqnarray}

Summarizing, we can therefore say that Eqs. (\ref{gen-struct}) express a
wide and rich variety of permutational identities, holding true at the
non-canonical, non-planar level, and arising from the antisymmetrization of $%
M\left( \geqslant 3\right) $-index tensors, e.g. in the $su(2)$ sector of
the 4-d. $\mathcal{N}=4$ SYM theory. The canonical, planar limit of the case
$M=3$, summed on all $k\in Z_{L}$, coincides with the previously known
relation (\ref{29}) \cite{Beisert:2003tq}.
\begin{equation*}
\end{equation*}

\subsection*{Appendix III}

\subsection*{Geometric \emph{''spin edge-differences''} \emph{Ans\"{a}tze}}

Defining the \emph{''spin edge-difference of order }$n$\emph{''} in the
following way:
\begin{equation}
\overrightarrow{\Delta S}_{k_{0}k_{1}...k_{n}}\equiv \overrightarrow{\Delta S%
}_{k_{0}k_{1}...k_{n-1}}-\overrightarrow{\Delta S}_{k_{1}...k_{n}}=%
\sum_{i=0}^{n}\left( -\right) ^{i}\binom{n}{i}\overrightarrow{S}_{k_{i}},
\label{37}
\end{equation}
and introducing the ($\gamma $-dependent) \emph{''splitting and joining
chain operator of order }$n$\emph{''} as in (\ref{38bis}), i.e. defining
\begin{equation}
\Sigma _{k_{0}k_{1}...k_{n}}\left( \gamma \right) \equiv \Sigma
_{k_{0}\gamma _{k_{1}}}\Sigma _{k_{1}\gamma _{k_{2}}}...\Sigma
_{k_{n-2}\gamma _{k_{n-1}}}\Sigma _{k_{n-1}\gamma _{k_{n}}},\text{ \ \ \ }%
\gamma \in S_{L},  \label{38}
\end{equation}
it is possible to rewrite Eqs. (\ref{34}) and (\ref{36}) respectively as
\begin{equation}
H_{2}=\frac{2}{N}\sum_{\left( k_{1},k_{2}\right) \in Z_{L}^{2},k_{1}\neq
k_{2}}\left[ \left( \overrightarrow{\Delta S}_{k_{1}k_{2}}\right) ^{2}-1%
\right] \Sigma _{k_{1}k_{2}}\left( \gamma \right) ,  \label{39}
\end{equation}
\begin{equation}
H_{4}=-\frac{2}{N^{2}}\sum_{\left( k_{1},k_{2},k_{3}\right) \in Z_{L}^{3},%
\text{ \ }k_{1}\neq k_{2}\neq k_{3}}\left( \overrightarrow{\Delta S}%
_{k_{1}k_{2}k_{3}}\right) ^{2}\Sigma _{k_{1}k_{2}k_{3}}\left( \gamma \right)
.  \label{40}
\end{equation}
Therefore, it is completely reasonable to formulate the following \emph{%
''spin edge-differences''} \emph{Ans\"{a}tze} for the $n$-loop, non-planar $%
su(2)$ spin-chain Hamiltonian ($n\in N$):
\begin{eqnarray}
H_{2n} &=&\left( -\right) ^{n+1}\frac{2}{N^{n}}\cdot  \label{41} \\
\cdot &&\sum_{\left( k_{0},k_{1},...,k_{n}\right) \in Z_{L}^{n+1},\text{ \ }%
k_{0}\neq k_{1}\neq ...\neq k_{n}}\left[ \left( \overrightarrow{\Delta S}%
_{k_{0}k_{1}...k_{n}}\right) ^{2}+\alpha _{n}\right] \Sigma
_{k_{0}k_{1}...k_{n}}\left( \gamma \right) ,  \notag
\end{eqnarray}
where $\alpha _{n}$ $\in Z$ is such that the spin part of $H_{2n}$ ia a
linear \emph{homogeneous} function of the quantities $\left( 1-P\right) $,
and $P$ is a permutation operator coming from the $n$\emph{-th order ''spin
edge-difference''} $\overrightarrow{\Delta S}_{k_{0}k_{1}...k_{n}}$. For
example, from Eqs. (\ref{39}) and (\ref{40}) we respectively get $\alpha
_{1}=-1$ and $\alpha _{2}=0$. Notice that the linking part $\Sigma
_{k_{0}k_{1}...k_{n}}\left( \gamma \right) $ of the Hamiltonian $H_{2n}$
will determine a symmetry under the inversion of the order of the $\left(
n+1\right) $-tet $\left( k_{0},k_{1},...,k_{n}\right) $, i.e. under the
exchange of site indices expressed by Eq. (\ref{exchange}).

As already stressed in \cite{Bellucci:2005ma1}, the \emph{''spin
edge-differences''} \emph{Ans\"{a}tze} (\ref{41}) for the explicit form of
the $n$-loop, non-planar $su(2)$ spin-chain Hamiltonian have a simple
meaning: i.e. $\Sigma _{k_{0}k_{1}...k_{n}}\left( \gamma \right) $
cyclically exchanges the incoming and outgoing ends of the chains adjacent
to the sites $k_{0},k_{1},...,k_{n}$. At the same time, the spin part
(modulo the constant $\alpha _{n}$) acts as the (square of the) discrete $n$%
-th derivative of the spin operatorial 3-vector along the new chain. In the
continuum limit, such \emph{Ans\"{a}tze} are compatible with the BMN
conjecture (\cite{Berenstein:2002jq}-\cite{Bellucci:2004rud}), yielding a
term $\sim \lambda ^{2n}\left( \partial ^{n}\phi \right) ^{2}$ as the $n$%
-loop contribution.

As it is evident from Eq. (\ref{41}), independently on the choice to have an
\emph{homogeneous} dependence of the spin part of $H_{2n}$ on terms $\left(
1-P\right) $, and thus, independently on the choice of the constant $\alpha
_{n}$ $\in Z$, the \emph{''spin edge-differences''} \emph{Ans\"{a}tze} imply
a \emph{linear} dependence of $H_{2n}$ on terms $\left( 1-P\right) $, and
thus, on the permutation operators $P$'s \emph{at any loop order} (i.e.
\emph{for any} $n\in N$). But this does \emph{not} seem to be the case.

Indeed, as already briefly mentioned in \cite{Bellucci:2005ma1}, the
attempts to check linearity in permutation operators, even by using the rich
set of permutational identities previously treated, fail already for $n=3$,
i.e. for the $3$-loop level.

Finally, it is reasonable to claim that the elegant and simply-meaning \emph{%
''spin edge-differences''} \emph{Ans\"{a}tze} (\ref{41}) for the explicit
form of the $n$-loop, non-planar $su(2)$ spin-chain Hamiltonian $H_{2n}$
unfortunately fail, because their main consequence on the structure (of the
spin part) of $H_{2n}$, namely the linearity in the permutation operators $P$%
's, fails to be checked also in the first non-trivial case, i.e. at the $3$%
-loop level.

\providecommand{\href}

\end{document}